\documentstyle[editedvolume,namedreferences]{crckapb}

\newcommand{\be}{\begin{equation}}
\newcommand{\ee}{\end{equation}}
\newcommand{\lab}[1]{\label{#1}}
\newcommand{\r}[1]{~(\ref{#1})}
\newcommand{\lsim}{{{}_{{}_<}^{~}\atop {}^{{}^\sim}_{~}}}
\newcommand{\gsim}{{{}_{{}_>}^{~}\atop {}^{{}^\sim}_{~}}}
\newcommand{\kms}{{\rm km}\,\rm s^{-1}}
\newcommand{\kpc}{{\rm kpc}}
\newcommand{\mas}{{\rm mas}}
\newcommand{\yr}{{\rm yr}}
\newcommand{\la}{{\leq}}
\newcommand{\feh}{\mbox{${\rm [Fe/H]}$ }}
\newcommand{\teff}{\mbox{$T_{\rm eff}$ }}
\newcommand{\aj}{{\it Astronomical Journal}}
\newcommand{\apj}{{\it Astrophysical Journal}}

\begin{opening}
\title{THE RR LYRAE DISTANCE SCALE}
\author{Piotr Popowski}
\author{Andrew Gould}
\institute{Department of Astronomy\\The Ohio State University\\
           174 W. 18th Avenue, Columbus, OH 43210, USA.}
\end{opening}

\runningtitle{THE RR LYRAE DISTANCE SCALE}
\begin{document}

\begin{abstract}
	We review seven methods of measuring the absolute magnitude $M_V$ of 
RR Lyrae stars in light of the Hipparcos mission and other recent developments.
We focus on identifying possible systematic errors and rank the methods by
relative immunity to such errors.  For the three most robust methods,
statistical parallax, trigonometric parallax, and cluster kinematics, we
find $M_V$ (at [Fe/H]$=-1.6$) of 
$0.77 \pm 0.13$, $0.71 \pm 0.15$, $0.67 \pm 0.10$.  These methods cluster
consistently around $0.71 \pm 0.07$.  We find that Baade-Wesselink and
theoretical models both yield a broad range of possible values (0.45--0.70 and 
0.45--0.65) due to systematic uncertainties in the temperature scale and
input physics.  Main-sequence fitting gives a much brighter $M_V=0.45\pm 0.04$
but this may be due to a difference in the metallicity scales of the cluster
giants and the calibrating subdwarfs.  White-dwarf cooling-sequence fitting 
gives $0.67\pm 0.13$ and is potentially very robust, but at present is too
new to be fully tested for systematics.  If the three most robust methods
are combined with Walker's mean measurement for 6 LMC clusters,
$V_{0,\rm LMC}=18.98\pm 0.03$
at [Fe/H]=$-1.9$, then $\mu_{\rm LMC} = 18.33\pm 0.08$.
\end{abstract}

\section{Introduction}
RR Lyrae stars are among the most popular local distance indicators.
Smith (1995) describes in detail their general properties.
One can measure the apparent magnitudes of RR Lyrae stars in a stellar system 
and infer their mean deredenned apparent magnitude $V_{0}$.
If the mean absolute magnitude of RR Lyrae stars $M_V(RR)$ at the system 
metallicity ${\rm [Fe/H]}$ is known, then the system's distance modulus $\mu$ 
is:
\be
\mu = V_{0} - M_V(RR). \lab{i1}
\ee
In the following sections, we will assume that $V_{0}$ can be measured 
accurately and concentrate on the $M_V$ determinations. $M_V$ can be calibrated
through field or globular cluster stars. There are compelling arguments
(e.g., Catelan 1998) against two distinct, environment-dependent $M_V$ scales 
(i.e., distance scales), and we will seek a universal absolute magnitude -- 
metallicity $M_V$ -- ${\rm [Fe/H]}$ relation:
\be
M_V(RR) = \alpha ({\rm [Fe/H]} + 1.6) + \beta. \lab{i2}
\ee

Most of the extragalactic distance scale is tied to the Large Magellanic
Cloud (LMC). The $M_V$ -- ${\rm [Fe/H]}$ relation from\r{i2} guarantees that
$\mu_{LMC}$ will be sensitive to the zero point $\beta$ and insensitive to the
slope $\alpha$. Therefore, we do not discuss many $\alpha$ determinations
(Ajhar et al. 1996; Fusi Pecci et al. 1996; Kov\'{a}cs \& Jurcsik 1996 etc.),
but concentrate on methods yielding $\beta$.
Note that $V_{0,{\rm LMC}} \approx 19.0$ (Walker 1992; Hazen \& Nemec 1992; Reid \& Freedman 1994) and so, from equation\r{i1} the division between the short and 
long distance scale occurs at $\beta \sim 0.55$, with fainter $M_V$ 
corresponding to the short distance scale.
Some methods determine $M_V(RR)$ directly using their
positions (trigonometric parallax), kinematics (statistical parallax) or
pulsational properties (Baade-Wesselink). Others (cluster kinematics, main
sequence and white dwarf fitting) establish globular cluster distances
and then, in a second step, $M_V(RR)$.
We group methods according to their mathematical description. Our, rather
incomplete, literature review serves mostly illustrative purposes, as we 
concentrate on the physical picture of the methods, emphasizing their 
strengths and weaknesses.

\section{Kinematic Methods}
	The distance to an ensemble of stars can be measured by
comparing their radial velocities ($\kms$) and 
proper motions ($\rm s^{-1}$) under the assumption
that these are due to (statistically) identical physical velocities.
In effect, one fixes the distance so that either the first moment 
of the population (bulk motion) as determined from the proper motions
is equal to first moment as determined from the radial velocities, or
so that the second moments (dispersions) are equal, or both.

	The great beauty of these methods is that the basic measurements
are of {\it dimensionless} quantities (redshift for radial velocities and
positions on the sky for proper motions) and therefore no assumptions about
the distance scale enter the determinations.  The major systematic uncertainty
(aside from concerns about the quality of the data) is that it may be difficult
to verify that the radial velocities and proper motions in fact arise from
the same physical velocity distribution.

	One may show that if the measurement errors are small compared to the
intrinsic dispersion of the population, then the fractional distance error
from equating the first moments is 
$\delta \eta/\eta = 1/[(A_1 N)^{1/2}\kappa\sin\theta]$ where $N$ is the total
number of stars, $\theta$ is the angular size of the system, $A_1$ is a
geometrical factor of order unity, and $\kappa$ is the ratio of bulk motion
to velocity dispersion.  Similarly the error from equating the second moments
is $\delta \eta/\eta = 1/(A_2 N)^{1/2}$ where $A_2$ is another
factor of order unity.  Hence, by combining the two methods, the
error is
\be
\biggl({\delta \eta\over \eta}\biggr)^2 = 
{1\over N[A_1(\kappa\sin\theta)^2 + A_2]}
\ee

	For nearby open clusters like the Hyades, $\kappa\sim 10^2$ and
$\theta\sim 10^{-1}$, so the first term in the denominator dominates,
and the distance is determined from the first moments.  This is sometimes
called the ``moving cluster'' method.  However, there are no RR Lyrae stars
in open clusters.  For globular clusters, $\kappa\sim 10^1$ and
$\theta\sim 10^{-3}$, so the second term dominates.  Hence, kinematic
distances to globular clusters are determined by matching velocity 
dispersions.  For
field RR Lyrae stars in the solar neighborhood, $\kappa\sim 2.1$ and 
$\sin\theta\sim 1$, so both the first and second moments are used to
determine the RR Lyrae distance scale.  This is called 
``statistical parallax''.  Statistical parallax and kinematic distance to
globular clusters have different
sources of systematic error, so we treat them separately.

\subsection{Statistical Parallax}

	Statistical parallax is reviewed thoroughly by Layden elsewhere
in this volume, so we give only a brief overview here.  In contrast to the
case of globular clusters, local field RR Lyrae stars are not at a common
distance.  Hence, before their radial velocities and proper motions can
be compared, the stars must be put on a common scale by taking account of their
dereddened apparent magnitudes.   That is, one can measure the 9
parameters describing the velocity ellipsoid (3 components of bulk motion
plus 6 independent components of the velocity-dispersion tensor) 
from radial
velocities alone.  On the other hand, if 
one {\it assumes} some arbitrary $M_V(RR)$,
for the RR Lyrae stars, then one can infer their distances from their
measured apparent magnitudes and estimated extinctions.  The distances and
proper motions yield the transverse velocities, and from these one can
again estimate the 9 parameters of the velocity ellipsoid.  One could
then adjust the assumed $M_V$ so that the velocity ellipsoid
from proper motions matched the velocity ellipsoid from radial velocities
as closely as possible.  In practice, one fits for all 10 parameters
(9 for the velocities plus $M_V$) simultaneously
using maximum likelihood.  The maximum likelihood approach was pioneered
by Clube \& Dawe (1980) and was subsequently applied by Hawley et al.\ (1986)
to the then best-available data set of 142 RRab stars.  They obtained
$M_V = 0.76 \pm 0.14$.  Note that the error is very close to the theoretical
minimum (for no measurement errors) 
$\sigma_{M_V} = (5/\ln 10)[(2 N/9)(6 +\kappa^2)]^{-1/2} = 0.12\,$mag
(Popowski \& Gould 1998a).  This
is because the measurement errors do not contribute significantly if they
are substantially below the velocity dispersion.

	As we discuss below, this result is fainter than
virtually all other estimates and much fainter than some.  Since the
method itself appears extremely robust, several workers have
invested substantial
effort to obtain the most reliable input data and to investigate whether
any unrecognized effects could be leading to systematic errors.  Layden (1994)
and Layden et al.\ (1996) put all existing data on a homogeneous system.  They
incorporated the proper motions from the new Lick NPM1 
(Klemola, Hanson, \& Jones 1993) survey which has smaller (and just as 
important, better-understood) errors than previous proper-motion studies.  
They used
the Lick catalog to calibrate the errors of the non-Lick proper motions.
They found that the diverse photometry sources were offset from one another
and put these on a common zero point.  Likewise, they put all the extinctions
on the Burstein \& Heiles (1984) system.  They found
$M_V = 0.71 \pm 0.12$ for 162 ``halo-3'' RRab stars with mean metallicity
$\langle{\rm [Fe/H]}\rangle = -1.61$.  Popowski \& Gould (1998a) developed
a formulation of maximum likelihood which permitted both new analytic
investigations and much more vigorous Monte Carlo investigations
of possible systematic effects.  These included a possible change in the
velocity ellipsoid with distance from the Galactic plane, rotation of the
local coordinate frames relative to the Sun's frame, and effects due to
the severe non-Gaussianity of the velocity distribution.  However, these 
effects
all proved negligible.  The most important previously overlooked effect that
they found was Malmquist bias, and they obtained 
$M_V = 0.76 \pm 0.12$ for the 162-star Layden et al.\ (1996) sample.

	Hipparcos has had two major impacts on the RR Lyrae statistical
parallax determination.  One is, of course, new and more precise 
proper-motion measurements.  The other, more surprisingly and indeed more
importantly, is better photometry.  Fernley et al.\ (1998a) fit Hipparcos
light curves to obtain new photometry for most RR Lyrae stars in the Hipparcos
catalog.  They combined these with Hipparcos proper motions and obtained
$M_V = 0.77 \pm 0.17$ for 84 ``halo'' ([Fe/H]$<-1.3$) stars with
$\langle{\rm [Fe/H]}\rangle = -1.66$, including 69 RRab's and 15 RRc's.
Fernley et al.\ (1998a) also used high-precision ground-based photometry
to show that their Hipparcos-based mean magnitudes were correct with very
small ($<0.02\,$mag) scatter.  
(Tsujimoto, Miyamoto, \& Yoshii 1998 conducted
a similar study of 99 Hipparcos ``halo'' stars and found
$M_V = 0.69\pm 0.10$ at $\left<\rm [Fe/H]\right>=-1.58$.  However, since their
quoted error is a factor 0.7 below the theoretical minimum, we conclude
that their analysis is incorrect.) Popowski \& Gould (1998b) used the Hipparcos
proper motions to check earlier catalogs and found that only the Lick
catalog is of sufficiently high quality to use.  Gould \& Popowski (1998) used
Fernley et al.'s (1998a) Hipparcos-based mean magnitudes to check 
Layden et al.'s (1996) systematization of previous heterogeneous photometry,
and found that it was 0.06 mag too bright.  They also incorporated the
new extinction map of Schlegel, Finkbeiner \& Davis (1998) based on COBE/IRAS
measurements of dust emission, and they eliminated a number of stars with
questionable extinctions and proper motions to obtain
$M_V = 0.77 \pm 0.13$ for 147 ``halo-3'' RRab stars with
$\langle{\rm [Fe/H]}\rangle = -1.60$.  

	Layden et al.\ (1996) noted that the velocity ellipsoid of
their solution is in good agreement with that of Beers \& 
Sommer-Larsen (1995) for metal-poor stars and took this as independent 
confirmation of the correctness of their results.  Popowski \& Gould (1998b)
and Gould \& Popowski (1998) directly incorporated the Beers \& Sommer-Larsen
stars into the analysis and obtained
$M_V = 0.80 \pm 0.11$ at $\langle{\rm [Fe/H]}\rangle = -1.71$ for a combined
sample based on 149 RRab Lyrae stars and 716 non-RR Lyrae stars.  

	At this point, essentially all
systematic errors have been removed from the RR Lyrae statistical parallax
determination.  The statistical errors in the above two solutions 
($M_V = 0.77 \pm 0.13$ at 
$\langle{\rm [Fe/H]}\rangle = -1.60$ or
$M_V = 0.80 \pm 0.11$ at $\langle{\rm [Fe/H]}\rangle = -1.71$)
should therefore be taken at face value.

\subsection{Kinematic Cluster Distances}

	If the distance to a cluster is known, the $M_V(RR)$
can be determined by subtracting the distance modulus from the dereddened
apparent magnitude of RR Lyrae stars in the cluster, or more generally from
the height of the zero age horizontal branch (ZAHB) at the color of the 
instability
strip.  Cluster distances can be determined kinematically by comparing
the dispersions of the radial velocities and proper motions.  The principle
is similar to statistical parallax, but there are two major practical
differences.  First, the proper motions are much smaller for clusters
$(\sim 10\,\kms/10\,\kpc= 0.2\,\mas\,\yr^{-1})$ than for nearby field
RR Lyrae stars $(\sim 200\,\kms/2\,\kpc= 20\,\mas\,\yr^{-1})$.  Second,
clusters are seen in projection, so the 3-space position of individual 
stars is unknown.  This introduces additional systematic effects that are
difficult to fully take into account.  

	Cudworth (1979) made the first such measurement, finding a distance
$d = 9.6\pm 2.6\,\kpc$ for M3 by comparing the proper-motion dispersion of 71
stars, $\sigma_\mu = 0.094\pm 0.021\,\mas\,\yr^{-1}$ with the radial-velocity
dispersion from Gunn \& Griffin (1979), $\sigma_r = 4.3\pm 0.7\,\kms$.
The measured rms dispersion was $\sigma_m =0.183\,\mas\,\yr^{-1}$ 
while the ``internal'' rms
measurement error was $\epsilon = 0.130\,\mas\,\yr^{-1}$.  Cudworth took
the ``external'' error  to be $f\epsilon$ and estimated 
$f=1.20\pm 0.12$.  Note that the measurement errors
were actually larger than the cluster dispersion 
$\sigma_\mu^2 = \sigma_m^2 - (f\epsilon)^2$ and that the correction factor
$f$ is therefore very important.  This factor had earlier been measured as
$f=1.1\pm 0.1$ by Cudworth \& Monet (1979) by comparing reductions of two
different plate sets of M13, one weak and one deep.  However, Cudworth (1979)
argued that this was only a lower limit on $f$ since the two sets were taken
with the same Yerkes telescope, over the same time interval, ca.\ 1900 to 
ca.\ 1975.  We note that a perfectly plausible $f\equiv 1$ would yield
$\sigma_\mu =  0.129\pm 0.023\,\mas\,\yr^{-1}$ and hence a cluster distance
of $d = 7.1\pm 1.7\,\kpc$.  

	Cudworth (1979) assumed that M3 is isotropic
because it looks circularly symmetric on the sky and
rotation is barely detectable if at all in the radial velocities.  Hence,
he simply divided the radial-velocity dispersion by the proper-motion to
obtain the distance.  However, if the cluster rotation axis were along the
line of sight, these observational characteristics would remain in tact,
but the radial dispersion would no longer be representative of the tangential
dispersion.  (Transverse rotation cannot be measured from the proper-motion
data because there are too few foreground stars and their dispersion is too
large to form a stable framework.)\ \  Thus, M3 exemplifies both the promise
and the problems of this technique.

	  Gunn \& Griffin (1979) introduced a major advance by 
incorporating Mitchie-King models (Mitchie 1963) into the analysis, and 
Lupton, Gunn, \& Griffin (1987) built on their experience when they
measured the distance to M13.  The Mitchie-King models are constrained by the
light profile, the radial-velocity map, and the mass function.  
Such models can take
account of rotation although the degeneracy noted in the previous
paragraph remains.  M13 is closer than M3 (6.5 vs.\ 9.6 kpc), 
the velocity dispersion is larger (6.5 vs.\ 4.3 $\kms$), and the number
of stars is larger (268 vs.\ 71).  Hence the distance errors are much smaller
(8\% vs.\ 27\%).  Lupton et al.\ (1987) do not say whether they incorporated
Cudworth \& Monet's (1979) estimate of $f=1.1$, or whether they used $f=1$.
In this case, the difference in the final result is 4\%, or 0.08 mag in the
height of the horizontal branch (HB).  Lupton et al.\ (1987) determined the
distance only for their best-fit model of the cluster and did not obtain
distances for a range of acceptable models which would have allowed them
to evaluate the systematic uncertainty due to their modeling procedure.

	Peterson \& Cudworth (1994) measured the distance to M22 by
equating the minor axis proper-motion dispersion with the radial-velocity
dispersion and found $d=2.6\pm 0.3\,\kpc$.  That is, they assumed that the
axis of the observed rotation is in the plane of the sky and, rather than
attempt to model the effect of this rotation on the observed proper motions
in the symmetry plane, simply ignored half the proper-motion data.  They
do not say whether they incorporated an $f$ factor, but for this case the
difference between $f=1$ and $f=1.1$ is only 0.03 mag.

	Rees (1996) has applied Gunn \& Griffin's (1979) Mitchie-King
technique to 8 clusters (47 Tuc, M5, M4, M92, N6397, M22, M15, and M2)
and combined the results with the distances obtained for M3 and M13 by
Cudworth (1979) and Lupton, Gunn \& Griffin (1987).  Subsequently,
Rees (1998 priv. comm.) rereduced the M22 data and derived a
distance, $d=11.26\pm 1.31\,\kpc$.  He also came to the conclusion that 
the N6397 proper motions were not reliable enough to use.  Two of the remaining
9 clusters have features which are not well represented by Mitchie-King models:
M15 has a central cusp, and 47 Tuc has differential rotation.  If these two
clusters are eliminated, the remaining seven have a mean zero-age HB (ZAHB) of
$M_V = 0.62 \pm 0.10$ mag at 
$\langle{\rm [Fe/H]}\rangle = -1.6$.  
  (Rees quotes an uncertainty of 0.05 mag, evidently
having renormalized the errors because $\chi^2 = 1.66$ for 6 degrees of
freedom.  However, since the errors are basically due to counting statistics
-- and so cannot have been overestimated -- 
the errors should not
be renormalized.)\ \  If M15 and 47 Tuc are re-included, 
$M_V = 0.58 \pm 0.08$.

	The details of Rees's (1996) work are not yet publicly available,
so it is impossible to give a complete assessment of the systematic
errors.  Clearly, if the proper motion errors have
been systematically underestimated (as suspected by Cudworth \& Monet 1979 and
Cudworth 1979), then the luminosity of the ZAHB has been underestimated for
each of the clusters.  This is a much smaller effect for nearer 
clusters which have smaller overall errors and so more statistical weight.
Alternatively, if there is no error underestimate but an underestimate
has been corrected for, then the luminosity of the ZAHB has been overestimated.
To date, no one has published a {\it set of distances} for a set of
acceptable Mitchie-King models, so there is no way to assess the systematic
error due to uncertainty in the cluster geometry.  At least until such tests
are conducted, clusters that are inconsistent with Mitchie-King models 
(i.e.\ M15 and 47 Tuc) should be excluded.

	Finally, there is bias introduced by binaries which increase the 
radial-velocity dispersion
but not the proper-motion dispersion (measured
on 100 yr time scales).  The correction for this previously unrecognized
bias makes $M_V$ fainter by
$\delta M_V=(2.5/\ln 10)f\langle (\gamma V_{\rm orb})^2\rangle/3\sigma^2\,$mag,
where $f$ is the binary fraction, $\sigma$ is the 1-dimensional dispersion
of the cluster, $V_{\rm orb}$ is the mean orbital velocity of the binaries,
$\gamma = M_s/(M_s+M_p)$, and $M_p$ and $M_s$ are the primary and secondary
masses.  Hut et al.\ (1992) conclude (primarily based
on Pryor et al.\ 1989) that for periods of 0.2 to 20 years, the binary
fraction of globular clusters is consistent with that of G stars as 
measured by Duquennoy \& Mayor (1991), i.e., 8\%.  Applying the above formula
to the Duquennoy \& Mayor (1991) distribution, we find
$\delta M_V = 0.05$.  Hence, our best estimate for this method is
$M_V=0.67\pm 0.10$ at $\langle{\rm [Fe/H]}\rangle = -1.6$.  We caution
that this error is only statistical and that the systematics are not fully 
understood.

\section{Trigonometric Parallax}

	Prior to Hipparcos, it was not possible to make a useful estimate
of the $M_V(RR)$ using trigonometric parallax.  Hipparcos measured
only one RR Lyrae star with reasonable precision, RR Lyrae itself.  This
yields $M_V=0.78\pm 0.29$ at [Fe/H]$=-1.39$ (Fernley et al.\ 1998a), still 
not precise enough to discriminate among various other estimates.

	However, Gratton (1998) has extended this approach by considering
all HB stars (not just RR Lyrae) in a magnitude limited sample, $V_0<9$.
This criterion yields 22 stars, including 10 blue HB stars, 3 RR Lyrae
stars, and 9 red HB stars.  Gratton (1998) obtains 
$M_V=0.69 \pm 0.10$ at $\langle{\rm Fe/H}\rangle =-1.41$ if all 22 stars
are included, or $M_V=0.60 \pm 0.12$ at $\langle{\rm Fe/H}\rangle =-1.51$ if 
one red HB star that he suspects of being a giant (not HB) star
is eliminated.  We first review and correct Gratton's statistical procedure,
and then discuss more carefully the problem of selection.

	Gratton (1998) begins by determining the {\it shape} of the HB
from observations of M5.  This yields $\delta M_V(B-V)$, the difference
between the magnitude at the instability strip and the magnitude of a star
at a particular $B-V$ color.  His model of the HB is therefore characterized
by a single parameter, $M_V$ at the instability strip.  From the observed
magnitude and color and the inferred reddening, he is then able to predict
the parallax as a function of this parameter, 
\be
\pi^*(M_V)= 10^{0.2\{[M_V - \delta M_V(B-V)] - [V - 3.1 E(B-V)] + 10\}}
\mas \lab{tp1}
\ee
He then forms the average of the difference between this quantity and the
observed parallax, $\pi$, weighted by the observational errors, and
determines $M_V$ by setting this quantity to zero
\be
0\equiv \Delta = \sum_i{\pi^*_i(M_V) - \pi_i\over \sigma_{\pi,i}^2} \lab{tp2}
\ee
Now, the proper procedure would be to form 
$\chi^2(M_V) = \sum_i (\pi^*_i(M_V) - \pi_i)^2/\sigma_{\pi,i}^2$ and then to
minimize it
with respect $M_V$.  Formally,
\be
0\equiv {\partial \chi^2(M_V)\over \partial M_V} = 
{\ln 10\over 2.5} \sum_i{\pi^*_i(M_V) - \pi_i\over \sigma_{\pi,i}^2}\pi^*_i(M_V) \lab{tp3}.
\ee
Comparing equations\r{tp2} and\r{tp3}, we see that Gratton (1998) has
in effect weighted the terms by their distance, which of course gives 
higher weight to the most poorly determined values.  Employing
equation\r{tp3} rather than\r{tp2}, we find $M_V=0.75\pm 0.09$ and
$M_V=0.64 \pm 0.12$ for the two cases.

	The red HB star that Gratton (1998) eliminated is suspicious only 
because it is significantly fainter than the HB.  Of course, since its
parallax is only $1.3\,\sigma$ larger than predicted, this could
be a normal statistical fluctuation.  There does not appear to be any
way to guard against giant-star contamination of the red HB sample without
biasing the sample by eliminating stars with above average, but normally
distributed, parallaxes.  Therefore the most prudent procedure is
to eliminate all red HB stars.  We then find
$M_V=0.68 \pm 0.14$ at $\langle{\rm Fe/H}\rangle =-1.62$.  Finally, we note
that account must be taken of the intrinsic scatter of the population.  For
homogeneous populations (e.g.\ globular clusters) the scatter in $M_V$
is $\sigma_{M_V}\sim 0.08$ mag.  
The upper limit for the scatter in a general field
population is $\sigma_{M_V}\sim 0.17$ mag, but a more plausible estimate is 
$\sigma_{M_V}\sim 0.14$ mag (Popowski \& Gould 1998a).  
We adopt that here and find $M_V=0.68\pm 0.15$.  This scatter introduces
a Malmquist bias.  One may show that if the parallax errors scale with flux
as $\sigma_\pi\propto F^{-\nu}$, then the Malmquist bias is 
$\delta M_V = (\ln 10/5)(1-\nu)\sigma_{M_V}^2 \sim 0.01\,$mag, where
$\nu\sim 0.2$ (appropriate for $6.5\la V \la 9$ for Hipparcos).
Our final estimate is therefore
$M_V=0.69 \pm 0.15$ at $\langle{\rm Fe/H}\rangle =-1.62$, where the
metallicity is on the Gratton scale, corresponding to $M_V=0.72 \pm 0.15$ at 
$\langle{\rm Fe/H}\rangle =-1.6$ on the Zinn \& West (1984) scale.

\section{Baade-Wesselink Method}
The goal of the Baade-Wesselink method is to determine
the equilibrium radius of a pulsating star through the combined analysis of 
its radial-velocity and light curves. The 
equilibrium radius and effective temperature, \teff, inferred from 
multicolor
photometry allows one to determine the luminosity, $L$.
The radial-velocity curve reflects the absolute change
in radius $R$ of a star, $\delta R$, due to its contraction or expansion
with velocity $V_{rad}(t)$. Consequently, during the time interval
$\delta t$:
\be
\delta R (t, t + \delta t) = \overline{V}_{rad}(t, t + \delta t) * \delta t, \lab{bw1}
\ee
where $\overline{V}_{rad}(t, t + \delta t)$ is the mean radial velocity in
a time interval $(t, t + \delta t)$.
The flux of a spherical 
blackbody at distance $d$ is given by:
\be
F = \frac{4\pi \sigma_S R^2 T_{\rm eff}^4}{d^2}. \lab{bw2}
\ee
where $\sigma_S$ is a constant.
Consequently, one can determine $R$
by:
\be
R = 2\delta R {\left( \frac{T_{\rm eff}^4(R)F(R+\delta R)}{T_{\rm eff}^4(R+\delta R)F(R)} - 1 \right)}^{-1} \lab{bw3}
\ee
In equation\r{bw3}, $\delta R $ is known from radial velocity curve 
[eq.\ \r{bw1}], the fluxes are observables, and the temperature at each phase
is estimated from colors based on stellar atmospheric models 
(e.g., Kurucz 1992).

Most of the results obtained before 1994 for field RR Lyrae stars
(Liu \& Janes 1990; Jones et al. 1992; Cacciari, Clementini \& Fernley 1992; 
Skillen et al. 1993) can be summarized by one equation:
\be
M_V = 0.70 + 0.21 \left({\rm [Fe/H]}+1.6\right), \lab{bw4}
\ee
resulting in $\mu_{\rm LMC} \approx 18.3$.

The Baade-Wesselink method is more involved and model-dependent
than geometric and kinematic determinations.
The simple physical picture of the method presented above gives an immediate
insight into its possible systematics.
First, $\overline{V}_{rad}$ in equation\r{bw1} is not the 
mean radial 
velocity that is measured from spectral lines. $\overline{V}_{rad}$ would
be equal to the measured velocity if the stellar photosphere were moving
straight toward or away from the observer. In fact it is moving radially inward
or outward from the center of a star and its direction forms some angle,
$\theta$, with the observer-star line of sight, which reduces the observed 
radial velocity by a factor of $\cos (\theta)$.
Additionally, a contribution of a given stellar surface patch to the
line intensity depends on limb darkening of the star and the relative depths
at which a given line and the stellar continuum form.
All these effects are usually parametrized by a factor ``$p$'' which is 
supposed
to convert the {\em measured} radial velocity to a {\em true} one 
(e.g., Fernley 1994).
Getting (1935) obtains $p = 1.41$ analytically using the Milne-Eddington limb 
darkening model. More sophisticated
investigations using model atmospheres usually produce slightly
lower values in the range 1.28 -- 1.39.
The factor $p$ is a function of pulsation phase, but
Fernley et al. (1989) showed for X Ari that using a varying value
of $p$ changes the derived $R$ by less than 1\%.
The factor $p$ also increases sharply with decreasing line strength 
(Karp 1975).
The uncertainty in $p$ does not influence the slope of the $M_V$ -- \feh
relation but does have a substantial impact on the zero point 
because $F \propto p^2$ [eqs.\r{bw3},\r{bw1} into\r{bw2}]. For small changes 
in $p$, the zero-point shift is
\be
\Delta \beta = - \frac{5}{\ln 10} \frac{\Delta p}{p} \lab{bw5}
\ee
Fernley (1994) argues that the original values of RR Lyrae $p$ coefficients
with a mean of $\sim 1.32$ (Skillen et al. 1993) should all be replaced with 
a universal value of $p = 1.38$.
Equation\r{bw5} then predicts $\Delta \beta = 0.09$, but
more detailed analysis (Fernley 1994) gives $\Delta \beta = 0.07$, leading 
to
\be
M_V = 0.63 + 0.21 \left({\rm [Fe/H]}+1.6\right). \lab{bw6}
\ee

Second, the unambiguous determination of the effective temperature from
stellar colors is still problematic. 
Temperature differences are likely to affect both the slope and zero point
of the  $M_V$ -- \feh relation.
\teff as obtained from optical colors 
($\lambda \lsim 800$ nm) 
are higher by 200-300 K than those inferred from $(V-K)$ colors.
Most observers prefer $(V-K)$ and usually argue that the existence
of $(U-B)$ excess in some pulsational phases is an indication
of atmospheric shocks. Therefore, optical colors are
likely to give an incorrect \teff.
However, McNamara (1997) claims that Preston \& Paczynski (1964) find
no correlation between ultraviolet excess and strength of hydrogen emission
lines suggesting that some other phenomenon is responsible for this effect.
He also argues that surface gravities as inferred from optical colors 
match much better the expectations from mass and radius values for RR Lyrae 
stars, whereas the surface gravities derived from $(V-K)$ are too small.
McNamara (1997) derives an $M_V$ -- \feh relation with a rather steep slope
and bright zero-point:
\be
M_V = 0.50 + 0.29 \left({\rm [Fe/H]}+1.6\right). \lab{bw7}
\ee
Which set of colors should be used to obtain the $M_V$ -- \feh
relation remains an open question.
Feast (1997) argues that even ``traditional'' Baade-Wesselink
analyses produce steep slopes. He claimes that $M_V$ should be treated as the 
independent variable rather than \feh. This is incorrect. If a regression
is done, the quantity with smaller relative errors should be the independent
variable. Maximum likelihood, which treats both errors equally is
the most rigorous approach and reproduces the shallow slope 
(Fernley et al. 1998b).
In addition, Feast's (1997) claim 
that we do not see the brightest RR Lyrae
stars due to a bias of the same sort of Malmquist bias is unlikely.
A more natural explanation for the lack of stars
with $M_V < 0.5$ is that they do not exist: the $M_V$ -- \feh relation
may simply flatten at low \feh as predicted by some theoretical models
(e.g. Caputo 1997).

Third, RR Lyrae stars are not perfect blackbodies and
bolometric corrections must be applied.
Even though RR Lyrae stars are rather hot and metal poor, so departures
from blackbody should be modest, the adopted bolometric
corrections actually differ by up to 0.1 mag (Sandage 1993 vs. Kurucz 1992). 

Similar investigations have been performed for globular cluster RR Lyrae
stars. Representative results obtained by Storm, Carney \& Latham (1994),
namely $M_V = 0.66$ at ${\rm [Fe/H]} = -1.4$ for M5 and $M_V = 0.45$ at 
${\rm [Fe/H]} = -2.24$ for M92, are slightly brighter than values predicted
by equation\r{bw4}.

\section{Theoretical Models of Horizontal Branch}

One can study the $M_V - {\rm [Fe/H]}$ relation
using a purely theoretical approach by constructing mock HBs in the $\teff - L$ plane and evolving them in time. The results depend sensitively
on the input physics (especially the opacities and equation of state: 
Castellani, Brocato \& Persimoni 1996), He abundance, and assumed core mass.
The treatment of convection is also one of the most essential ingredients.
To compare theoretical HBs to HBs observed
in globular clusters one must transform from the theoretical $\teff - L$
plane to the $M_V$ -- color plane.
This step requires accurate atmosphere models (e.g. Kurucz 1992). 
Then one can investigate HB morphology, how HB appearance depends
on different parameters (e.g., age) etc. One can also establish how 
$M_V({\rm ZAHB})$ at the instability strip scales with the ${\rm [Fe/H]}$ of a 
given stellar ensemble and how this relation
depends on the evolutionary stage of the population.
With some (even conservative) observational constraints taken into account, 
the model predictions become quite robust.
Another approach involves the analysis of the pulsational properties
of RR Lyrae stars in the spirit of the pioneering work by van Albada \& Baker
(1971). Pulsational models predict the relations between $T_{\rm eff}$,
${\rm [Fe/H]}$, luminosity, $L$, period, $P$, and mass, $M$, of the stars; 
e.g.,
\be
\log{P} = 11.497 + 0.84 \log{L} -0.68 \log{M} -3.48 \log{T_{\rm eff}}
\ee
They also enable one to investigate the location of blue and red, fundamental
and first overtone instability strip boundaries (Bono et al. 1995).
Which regions of the instability strip are populated by RRab and RRc stars
correlate with the Oosterhoff type of the cluster and gives insight
into the source of the period-shift effect (Sandage 1981,1982; 
Bono et al. 1995)
which has been used to determine the $M_V$ -- ${\rm [Fe/H]}$ relation.
The most robust constraints on the  $M_V$ -- ${\rm [Fe/H]}$ relation
come from combining the pulsational and evolutionary scenarios.

The predictions of theoretical evolutionary models are sensitive to
every change in input physics. For example,
recent improvements in opacities and the equation of state led to an increase
in HB luminosities by $\sim 0.1$ mag. Gratton et al. (1997)
report an unpublished  $M_V$ -- ${\rm [Fe/H]}$ relation by
VandenBerg et al. (1998):
\be
M_V(ZAHB) = 0.19 ({\rm [Fe/H]} + 1.6) + 0.66
\ee
Salaris et al. (1997) obtain a  slightly brighter result with a similar slope:
\be
M_V(ZAHB) = 0.21 ({\rm [Fe/H]} + 1.6) + 0.57
\ee
However HB luminosities are very sensitive to the He core mass. Mazzittelli
et al. (1995) suggested that core masses should be increased by $0.01 M_{\odot}$ relative to older estimates. Models with heavier cores have
slightly steeper and brighter HBs, e.g. Caloi et al.'s (1997)
non-linear $M_V$ -- ${\rm [Fe/H]}$ relation which can be approximately
characterized by
\be
M_V(ZAHB) = 0.26 ({\rm [Fe/H]} + 1.6) + 0.49
\ee
Caputo et al.'s (1993) and slightly adjusted Caputo (1997) HB models predict 
a break in slope which occurs near ${\rm [Fe/H]} = -1.3$ if RR Lyrae stars 
have scaled solar abundance of $\alpha$ elements, and near 
${\rm [Fe/H]} = -1.6$ if $\alpha$ elements are enriched by a factor of 2--3. 
These models predict a steep slope of 0.3 for metal-rich RR Lyrae stars and a
shallow slope of 0.19 for metal-poor stars. The zero point favors the 
short distance scale with $\mu_{LMC} \approx 18.4$.
Caputo (1997) checks her synthetic HB evolutionary models
proving that they satisfy the constraints produced by convective pulsating 
models with a range of masses and He abundance $Y=0.24$.
Such models allow one to predict the borders of the RR Lyrae distribution
in the $P$ -- $M_V$ plane.
Moreover, Caputo (1997) concludes that adopting the ``long distance scale'' 
with a steep slope for all metallicities ($M_V = 0.3 {\rm [Fe/H]} + 0.94$) 
results in RRc stars falling in the hot stable region, i.e. outside the 
instability strip.
On the other hand, the calibrations with a shallow slope are in agreement
with observations only if the zero point suggested by Fernley (1994) is adopted
instead of a fainter value.

Lee, Demarque \& Zinn (1990) and Caputo et al. (1993) analyzed HB
evolutionary tracks showing explicitly how RR Lyrae luminosity
depends on evolutionary stage and that blue, less-massive
HB stars cross instability strip at significantly higher $L$
than the ZAHB. Therefore the average RR Lyrae luminosity
will depend on which part of the HB they come from and so what is the 
HB morphology. Particular $M_V - {\rm [Fe/H]}$ relations will therefore
in general depend
on the HB morphology of the cluster. If the cluster contains $B$ blue HB stars,
$V$ RR Lyrae stars, and $R$ red HB stars, then its morphology index (Lee 1989),
$(B-R)/(B+V+R)$, is a number in the interval $(-1,1)$.
Morphologies below -0.8 (47 Tuc) and above 0.8 (M92)  indicate very red and 
very blue HBs and, according to Caputo (1997), are likely
to complicate the picture.
However, even if one restricts oneself to intermediate morphologies,
the uncertainties of HB models (at least 0.1 mag/dex
in slope and 0.2 mag in zero point) prohibit any precise conclusions
about the theoretically-preferred $M_V$ -- ${\rm [Fe/H]}$ relation.

There is also another theoretical path leading to a determination of $M_V$.
It progresses through the analysis of double mode RR Lyrae stars which is 
rather complex and is beyond the scope of this paper. Representative results
(Alcock et al. 1997, Kov\'{a}cs \& Walker 1998) tend to give a bright $M_V$.

\section{Main or White-Dwarf Sequence Fitting to Globular Clusters}

The main sequence (MS) or white-dwarf (WD) cooling-sequence fitting techniques
give the 
distance to a cluster by matching the cluster sequence of subdwarfs or WDs
to a corresponding sequence in $M_V$ -- dereddened color 
plane as determined from calibrating stars in the solar neighborhood.
The $M_V$ of the HB can be determined from the equation:
{\small
\begin{eqnarray}
M_V(RR) & = & V(RR) - A_V(RR) - \left[ V(CS) - M_V(LC) - A_V(CS) + \sum_i \frac{d M_V}{d \gamma_i} \delta \gamma_i \right] \nonumber \\
&\approx & V(RR) - \left[V(CS) - M_V(LC) + \sum_i \frac{d M_V}{d \gamma_i} \delta \gamma_i \right], \label{sf1}
\end{eqnarray}
}
\makebox[-1em]{}
where $V(RR)$ and $V(CS)$ are the apparent magnitudes of RR Lyrae stars in a 
cluster and the cluster sequence, respectively, $A_V(RR)$ and $A_V(CS)$ are 
the visual extinctions toward them, and $M_V(LC)$ is the
$M_V$ of the local calibrators at a given dereddened color.
Finally, $\sum_i \frac{d M_V}{d \gamma_i} \delta \gamma_i$ corrects for
the fact that for cluster stars it is not always possible to find local
calibrators with the same characteristics (parameter $\gamma_i$) or just
for parameter uncertainty of order of $\delta \gamma_i$. For MS
fitting the most essential parameters are metallicity and color 
(e.g., $(B-V)$), for WD fitting --- mass and color.
Note that if the cluster sequence stars experience the same extinction
as RR Lyrae star, then in equation\r{sf1}, $A_V(RR)$ and $A_V(CS)$ cancel out.
In practice it is enough if MS or WD sequence stars are drawn from
the same region of the cluster as RR Lyrae stars.
There are a few crucial criteria that must be met to make the fitting method
work properly:
\begin{enumerate}
\item accurate distances to the local calibrators must be known. Only when
one knows the distances is it possible to construct MS or WD
curves in the $M_V$ -- dereddened color plane. In principle
one should also know the reddenings to individual stars, but for the 
calibrating stars, most of which sit in the Local Bubble, this is not a major 
concern.
\item if the results of the sequence fitting are sensitive to some parameter,
numerous stars in different ranges of this parameter should be available.
Metallicity is a key parameter for the MS fitting because the $M_V$ of the MS
at the given color is a function of metallicity, especially
at the high metallicity end. One wants to avoid using theoretical models
to shift MS defined at one metallicity to get the $M_V$ for another.
\item one must measure the parameters of the cluster on the same scale
as applied to the local calibrators, e.g.,  
the metallicity of a cluster should be measured on the same
scale as the one used for local subdwarfs. It is necessary to match the 
appropriate
local sequence to the one observed in a cluster.
One possible cause of a systematic error in the MS fitting
distances is that the metallicities of the local subdwarfs might be
on a different scale from those of the clusters (determined
from giants). Even if the procedure to obtain each type of metallicity
is completely uniform, the conclusions depend on the atmospheric models,
which are completely different for subdwarfs and giants.
Specifically, if the subdwarf metallicities were too low
(or the giant metallicities too high) then the cluster distance
and the HB luminosities would be overestimated. King et al. (1998) found
intriguing evidence of a possible misalignment of this sort. They measured the 
metallicities of M92 subgiants (unfortunately not subdwarfs, but with
higher gravity than giants), and obtained metallicities up to half a dex
lower than those of M92 giants.
Pinnsoneault (1998, priv. comm.) finds a metallicity of M5 from $BVI$ colors
of MS stars. His value is $\sim 0.3$ dex more metal-poor than Gratton et al.'s
(1997) giant-based determination.
\item reddening corrections to the color must be known 
very accurately.
Reddening changes the conclusions about the $M_V$
of MS or WD stars in two major ways. 
First, through the extinction $A_V$, it 
directly dims the light of the observed stars shifting the cluster sequence 
vertically on the color magnitude diagram. However, this effect almost 
entirely cancels the $A_V$ correction when one determines 
the $M_V$ of the HB
--- both types of stars are extincted by approximately the same amount.
And second, uncertainty in color, $\delta (X-Y)$, results in the shift 
in $M_V$ of order of $\delta M_V \sim \frac{d M_V}{d (X-Y)}
\cdot \delta (X-Y)$. This effect is particularly dangerous if the $M_V$ --  
color relation is very steep (e.g., for $(B-V)$ color of MS stars, 
$\delta M_V \sim -5 \cdot \delta (B-V) $).
\end{enumerate}

\subsection{Main Sequence Fitting}

We stress at the outset that the MS fitting 
technique is very sound when applied to nearby open clusters.
For a given metallicity there are plenty of local calibrating dwarfs that 
define well constrained sequences. Because they are numerous there is
no problem with choosing a sample with very accurate parallaxes and so small 
errors in $M_V$.
On the cluster side - because open clusters are close, the spectroscopy of
the MS stars is straight forward. Therefore, the clusters and local calibrators
are easily put on the same metallicity scale. 
Globular clusters pose much more severe problems. Their metallicities are
generally very low and, consequently, there are few local subdwarfs
that can define metal-weak MS's. The local subsample is collected
from a larger region, which translates into larger parallax errors.
Therefore some of the distances to individual stars require a substantial
Lutz-Kelker type correction (Lutz \& Kelker 1973; Hanson 1979; Smith 1987).
Additionally, globular cluster are far away and their MS's
are too faint for high-resolution spectroscopy. Therefore the cluster
metallicities come from giants and not from subdwarfs.
Reddening to globular clusters is usually measured from integrated
photometry (e.g., Zinn 1980; Reed, Hesser, \& Shawl 1988), and potentially 
more accurate Str\"{o}mgren photometry is generally available only for 
southern clusters.
Moreover, for unknown reasons, Str\"{o}mgren-photometry-based reddenings 
are systematically smaller than other determinations. As we mentioned above,
reddening uncertainties can dramatically affect the distance
modulus.

Reid (1997) uses 15 subdwarfs with Hipparcos parallaxes accurate to at least 
12\% to define local subdwarf sequences of known $M_V$.
He separates the calibrating stars into two metallicity bins: intermediate
with  ${\rm [Fe/H]} \sim -1.4$ on the Zinn \& West (1984) scale and metal-poor
with
${\rm [Fe/H]} \sim -2.1$. That is, Reid (1997) does not apply any theoretical
transformations to obtain the $M_V$ of MS stars
at a given color. He reports that observational sequences are much less
sensitive to metallicity than predicted by theoretical isochrones.
However, this effect may be spurious because Reid (1997) uses Carney (1994) 
metallicities that are based on low-resolution spectra and therefore are 
likely to show more metallicity spread than is really present.
He derives new distances to 7 globular clusters: M5, NGC 6752, M13,
M15, M92, M30, and M68 obtaining distance moduli of 14.45, 13.17, 14.48,
15.38, 14.93, 14.95, and 15.29, respectively, with formal uncertainties
of 0.1 mag. For the last four globulars which are metal poor, these new
distance moduli are about 0.3 mags higher than the previous standard values.
This leads to intrinsically bright HB's (e.g., 
$M_V = 0.44 \pm 0.08 $ at ${\rm [Fe/H]} = -1.6$)
with an extremely steep $M_V$ --- ${\rm [Fe/H]}$ slope of 0.5.

Gratton et al. (1997) have a bigger local sample, but still believe that they 
do not have enough calibrating subdwarfs to make completely model independent 
fits to globular clusters. Instead, using calibrating subdwarfs, they determine
the relation between $(B-V)$ color for unevolved MS stars
as a function of metallicity. Then they use this relation to find the 
location of the MS on $M_V$ -- color diagram.
Gratton et al. (1997) determine distances to 9 globular cluster:
6 coinciding with Reid (1997) [except M15], plus 47 Tuc, NGC 288, and 
NGC 362.
Gratton et al. (1997) distances are generally slightly shorter than 
the ones by Reid (1997), but they seem to be systematically shorter by 0.2 mag
for the three of four metal poor clusters: M30, M68, and M92.
Gratton et al. (1997) use the metallicity scale of Carreta
\& Gratton (1997) which is more metal-rich than Zinn \& West (1984) scale
adopted by Reid. All these
modifications lead to,
\be
M_V(RR) = (0.18 \pm 0.09)({\rm [Fe/H]} +1.6) + (0.45 \pm 0.04). \lab{sf2}
\ee
where we report not the result from original paper but rather the corrected
relation as reported by Gratton (1998).
Equation\r{sf2} implies $\mu_{LMC} \approx 18.60 \pm 0.07$.

Studies by Reid (1997) and Gratton et al. (1997) devote a lot of effort to 
correcting for possible systematic
effects like Lutz-Kelker corrections, binary contamination, and errors
in reddening determinations.
Both studies use very similar input data. Nevertheless, they attribute their 
corrections to the pre-Hipparcos MS fitting results to two different sources. 
Gratton et al. (1997) argue that smaller parallaxes to calibrating subdwarfs 
by themselves account for 0.2 mag larger distances to globular 
clusters. Reid (1997) sees the same trend (Hipparcos magnitudes smaller than 
the ground-based measurements), but attributes this change mostly to the lack 
of sensitivity of the subdwarf luminosity to metallicity for stars with 
$-1.5 \gsim {\rm [Fe/H]} \gsim -2.0$ (which is very unlikely from a theoretical
point of view).

There is, however, some evidence that those two comprehensive studies are
not the ultimate in MS fitting analysis following the Hipparcos breakthrough. 
Pont et al. (1998) opt for using
evolved stars and binaries in their analysis, which dramatically increases
the number of calibrating stars. They find $\mu_{\rm M92}=
14.67$, resulting in $\beta \sim 0.6$ [compare to eq.\r{sf2}]. Using only 
unevolved calibrating stars, Pinnsoneault (1998, priv. comm.) argues for a 
similar $\beta$ from multicolor analysis of M5 (see also above). 

\subsection{White Dwarf Sequence Fitting}

The main theoretical difficulty one encounters using WD cooling 
sequences is the dependence of their magnitude on WD mass. Renzini et 
al. (1996) list four independent observations that constrain WD 
masses: luminosities of the red giant branch tip, the HB, the 
asymptotic giant branch (AGB) termination and the post-AGB stars.
All of them seem to indicate that masses of WD's in globular clusters
($M_{CWD}$) are independent of the host cluster metallicity (excellent!) and 
approximately
equal to $M_{CWD} = 0.53 \pm 0.02 M_{\odot}$ (Renzini \& Fusi Pecci 1988). 
(Such a narrow range of masses has never been either confirmed or 
falsified observationally, e.g., Cool, Piotto \& King 1996 claim that their 
data for NGC 6397 are consistent with the wider range $M_{CWD} = 0.55 \pm 0.05
M_{\odot}$, although they allow that all the spread may originate from the 
observational errors.) 
Wood's (1995) theoretical models
suggest that the $M_V$ of the WDs
scales with the WD mass as
\be 
\delta M_V \sim 2.4 \delta M_{WD} \lab{wd1}
\ee 
Therefore, a serious misestimate of the white dwarf $M_V$'s
requires a very severe violation of the WD mass constraints coming 
from a wide variety of theoretical predictions. According to Renzini 
\& Fusi Pecci (1988) the $M_V$ uncertainty produced by differences of
WD mass in different globular clusters can be of order of 0.05; if 
a rather conservative range $M_{CWD} = 0.55 \pm 0.05 M_{\odot}$ for the mass 
of the cluster WDs is adopted, $\delta M_V$ can reach 0.12. 
Nevertheless, all observers agree that within a given system, the distribution
of masses is very narrow. This tendency is confirmed by the local white dwarfs
used to calibrate the $M_V$ of the WD cooling sequence (Bergeron, 
Saffer, \& Liebert 1992; Bragaglia, Renzini, \& Bergeron 1995).
For the
solar neighborhood the mean and dispersion of the WD masses are 
$\left< M_{WD} \right> = 0.59 \pm 0.1$, but this does not  contradict the 
Renzini \& Fusi Pecci (1988) claim of a very narrow mass range in globular 
clusters, where WD populations come from low-mass progenitors only. 

There may be also some photometric errors leading to misestimates of $(B-V)$, 
but they are not likely to shift $M_V$ by more than 0.1 mag 
(in a random direction). Moreover, WDs have practically metal-free 
atmospheres that come in two varieties: pure H (DA) constituting 
the great majority and pure 
He (DB), which makes modeling very straight forward. They are more
abundant than subdwarfs and consequently it is much easier to collect
a large calibrating sample with accurate trigonometric parallaxes.
There is, however, one main observational disadvantage of using WD 
cooling sequences, namely they are at least 10 mag fainter than MS stars 
of the same temperature.
For the closest globular clusters, WDs
have $V \gsim 23$ (De Marchi, Paresce, \& Romaniello 1995; Richer et al. 1995,
1997; Cool et al. 1996) and so HST is the best tool to acquire 
photometric data. 

Even though WD sequences have been observed
for at least 6 globular clusters (47 Tuc, $\omega$ Cen, NGC 6397, M15, M4, 
NGC 6752), the distance to only one globular cluster was determined by the
WD fitting technique (Renzini et al. 1996 for NGC 6752), one result
was mentioned parenthetically (Richer et al. 1997 for M4) and there exist 
persistent rumors about one not yet published (Renzini 1998 for 47 Tuc).
Renzini et al. (1996) match the WD sequences (when DA is matched,
DB falls in the right place) of NGC 6752 with the local WD sequence and 
find $\mu_{\small N6752} = 13.05$. When combined with the reddening
of the cluster and apparent magnitude of the HB from 
Gratton et al. (1997) as originally observed by Penny \& Dickens (1986), this 
yields $M_V(RR) = 0.59 $ at ${\rm [Fe/H]} = -1.4$.
Renzini (1998) has also recently reported a distance modulus to 47 Tuc
that is about 0.5 mag shorter than the distance obtained through
MS fitting and would imply  $M_V(RR) = 0.95$ at ${\rm [Fe/H]} = -0.7$. 
Richer et al. (1997) mention that if they use Renzini \& Fusi Pecci (1988)
cluster white dwarf mass estimate of $M_{WD} = 0.53 \pm 0.02 M_{\odot}$, 
their (dereddened) distance modulus to M4 will be 11.09. 
Adjusting the HB $M_V$ of Rees (1996)
to match the Richer et al. (1997) distance modulus to M4, one obtains 
$M_V=0.75$ at ${\rm [Fe/H]} = -1.3$. 
Due to the very small number of analyzed clusters, the WD cooling
sequence technique must be interpreted cautiosly, but 
nevertheless seems to give some support to the short RR Lyrae distance scale.
For the 3 clusters (and fixing the slope of $M_V$ -- ${\rm [Fe/H]}$ relation
at 0.2), we find $M_V(RR) = 0.67 \pm 0.13$ at ${\rm [Fe/H]} = -1.6$.
 
\section{Summary}

\begin{table}[htb]
\begin{center}
\caption{Comparison between different methods to determine RR Lyrae
absolute magnitudes}
\begin{tabular}{lcclc}
\hline
& Absolute &&& Potential \\
Method & magnitude & Grade & Main problems & future \\
& at ${\rm [Fe/H]}=-1.6$ &&& usefulness \\
\hline
Statistical parallax & $0.77 \pm 0.13$  & A & --- & \makebox[0.8em]{}A$-$ \\
Trigonometric parallax & $0.71 \pm 0.15$ & \makebox[0.8em]{}A$-$ & done for non-RR Lyrae stars & \makebox[0.8em]{}A$+$ \\
&&& small number statistics & \\
Cluster kinematics & $0.67 \pm 0.10$ & \makebox[0.8em]{}B$+$ & modeling of rotation & \makebox[0.8em]{}A$-$ \\
&&& uncertain density profile & \\
&&& proper motions & \\
Baade-Wesselink & 0.45--0.70 & B & temperature scale & B \\
&&& {\it p} factor, bolometric corrections & \\
Theoretical models & 0.45--0.65 & \makebox[0.8em]{}B$-$ & input physics & A \\
Main sequence fitting & $0.45 \pm 0.04$ & C & metallicity scale & B \\
&&& reddening uncertainties & \\
&&& small number of calibrators & \\
White dwarf fitting & $0.67 \pm 0.13$ & C & WD masses from theory & A \\
&&& reddening uncertainties & \\
\hline
\end{tabular}
\end{center}
\end{table}

The most crucial outcome of the Hipparcos mission is the realization
that astrometry is at the moment the only fully reliable way to measure 
distances in the local Universe. However, instead of removing a long-standing 
discrepancy among existing distance scales, Hipparcos results led to numerous
analyses which have strengthened old conflicts and created new ones.
Table 1 summarizes the results reported in this review. Note that the errors
quoted are statistical only. We have ranked and graded (A -- very good; B -- good; C -- acceptable) the methods by
our judgement of their susceptibility to systematic effects and also included
our grade of their potential reliability with forseeable improvement in the
data. Of the methods listed only 3 (statistical parallax, trigonometric
parallax and MS fitting) have been affected by Hipparcos.

This work was supported in part by grant AST 97-27520 from the NSF.

\end{document}